\documentclass[preprint,12pt]{elsarticle}
\usepackage{xcolor}

\usepackage{amssymb}

\journal{Journal of Theoretical Biology}

\begin{document}

\begin{frontmatter}

		\title{A Prerequisite for Life}

\author{S\o ren Toxvaerd }

\address{ Department
 of Science and Environment, Roskilde University, Postbox 260, DK-4000 Roskilde, Denmark}

\begin{abstract}
The complex physicochemical structures and chemical reactions in living organism have some 
common features: (1) The life processes take place in the  cytosol  in the cells,
which, from a physicochemical point of view
	is an emulsion of biomolecules in a dilute aqueous suspension. (2) All living systems are homochiral with
	respect to the units of amino acids and carbohydrates, but (some) proteins are chiral unstable
	in the cytosol. (3) And  living organism are mortal.
These three common features together give a  prerequisite for the prebiotic self-assembly at the start of
the Abiogenesis. Here we argue , that it all together indicates, that the
prebiotic self-assembly of structures and reactions took place in a more saline environment, whereby the homochirality
of proteins not only could be obtained, but also preserved. A more saline environment for the prebiotic self-assembly
	of organic molecules and establishment of biochemical reactions could have been the hydrothermal vents.
\end{abstract}


\end{frontmatter}
	\section{Introduction}
Numerous  articles  deal with  the  environmental and physicochemical conditions for -,
and with the chemical reactions  at the establishment of
life on planet Earth, or somewhere else in the universe
\cite{Bada2004,Lane2010,Huber,Hud,Pascal,Stueken,Sousa,Sutherland,Spitzer,Walker,Branscomb}.
Many of the suggestions are inspired by Darwin, and  link these conditions and
 reactions with the  chemical composition  and reactions in LUCA, our
last  universal common ancestor \cite{Velasco2018,Higashi}. It is, however, not clear whether LUCA
is a simple Archaea or Bacteria, or whether it is a more primitive organism,
a Progenote \cite{WoeseFox1977,Woese1987,DiGiulio2011,Weiss2016,DiGiulio2018}.
For a recent review see \cite{Weiss2018}.
But although LUCA, superficially looks  rather  simple, it is in fact
very complicated and highly structured  \cite{Weiss2016,Weiss2018}, and with all the fundamental preconditions for our life.
LUCA has  a genetics  and a metabolism controlled by  enzymes in a cell with ion channels and active transport.

Our first ancestor is at least 3.5 billion years (Ga) old
\cite{Djokic2017,Schopf2018}, and there is indirect signs of life even before 3.5 Ga \cite{Tashiro2017,Retallack2019}.
The earth is $\approx$ 4.6 Ga old and there is good evidence for, that the oceans were established shortly
after, maybe after few hundred million  years \cite{Harrison,Sleep}. The prebiotic time
defined as the time interval, from where the physicochemical conditions allowed
for an inorganic and organic self-assembly of bio-materials to the time,
where LUCA appeared, is most likely   millions of years.
This article deals with  the start of the  Abiogenesis: the spontaneous chemical self-assembly of the organic
building units and the establishment of  reactions
in a prebiotic environment.

There are many universal properties of all the living species, which might guide us in establishing a  prerequisite  for life.
On one hand  living species are very complex, even in the details.
Take for instance the metabolism, the genetics or our immune system, or...
But on the other hand they have some common features. All living species are from a thermodynamic point of view open \ ``driven" system far
from equilibrium. The biochemical reactions in the cells
are in a diluted aqueous (cytosol) solution (1),  the  units of the peptides and carbohydrates are all  homochiral units (2),
and    living species are mortal (3).
 There are many other common features of biosystems. Beside of the
metabolism and the  genetics with the DNA and RNA, also 
e.g. the selfreplication and the cells with membranes. And with respect to mortality
there exists some counter-examples of living organisms with a remarkable
resistance to aging. For instance  some   primitive  organisms, \textit{Hydra vulgaris},
show no sign of mortality for a period of four years \cite{Martinez1998}, and some jellyfish
perform life cycle reversal \cite{He2015}. Another example is the stability of the
enzyme telomerase and the telomere. Together they are responsible for the
unlimited proliferation of almost all cancer cells \cite{Cong2002}.

With respect to the
necessity of homochirality of the units of amino acids and carbohydrates, it  is in general not debated in
articles, which deal with the prerequisites for life. But
 here  we argue,  that the
three common properties: the composition of the cytosol, the instability of homochirality in enzymes
and mortal instability of living systems, are  connected,
and that they together also is a key to determine  a prerequisite for  our life: prebiotic self-assembly of
homochiral enzymes in a saline aqueous environment.

	\section{The aqueous cytosol solution in the cells, homochirality and life}

All the cells  in living organisms  are soft condensed matter, limited by  cell membranes and the cells contain an aqueous (cytosol) suspension of
organic molecules.
The cytosol can be characterized as a diluted aqueous solution of cat- and anions at a total 
concentration (exclusive amino acids) of $\approx$ 0.15 M of $[$Na$^+]\approx$ 0.01 M, $[$K$^+]\approx$ 0.10 M, $[$HCO$̣_3^{-}]\approx$ 0.008 M,
and other ions, Cl$^-$, Ca$^{++}$, Mg$^{++}$,..,  with smaller concentrations. Despite
the small ionic concentrations in the cell interior, the cytostol with a complex 
cytoplasm with water networks and hyperstructures \cite{Shepherd2006,Norris2007}, departs physicochemical
from the condition of an ideal aqueous solution \cite{Luby-Phelps2000}.

LUCA is at least 3.5 billion years (Ga) old, and it is therefore natural to describe  how the 
earth was,  and how water appeared, at that time.
The earth is $\approx$ 4.6 Ga old and the Hadean eon is the period
from 4.6 Ga to 4.0 Ga,
followed by the Achaean eon from 4.0 Ga to 2.5 Ga.
There is  evidence for, that the ocean(s) were established
 relatively shortly after 4.6 Ga, maybe after one hundred million  years \cite{Harrison,Sleep}.
The estimates of the physicochemical state of the water in the oceans after the establishment varies
from prediction of an extreme salty and  acidic Hadean ocean \cite{Maruyama,Aoki} to an Archean ocean,
with a salt content as in the present oceans \cite{Marty}. Today the oceans  contain a sodium
concentration of $[$Na$^+] \approx$ 0.47 M and  $[$K$^+] \approx 10^{-2}$ M. The difference between the concentrations
of sodium and potassium outside and inside the cell is vital and maintained by the sodium pump.

Living systems consist of peptides with units of L-amino acids and  D-carbohydrates. The amino acids have, however,
an active isomerization kinetics  and will racemize in an aqueous solution  \cite{Bada}. 
The chirality of a peptide in a cell is also unstable \cite{Fujii2018}. But from
computer simulations and thermodynamic investigations of aqueous suspensions of peptide-like molecules
at different water activities  one finds, that although
peptides lose their homochirality at high water activity, they are stable and maintain homochirality
at a lower water activity (saline solutions) \cite{Toxvaerd2017}, in accordance with the observation of the stability of homochirality in a peptide.
 The physicochemical explanation for this behaviour is, that the
 compact  $\alpha$-helix  structure, predicted by Pauling $\textit{et al.}$ \cite{Pauling1,Pauling2},
 ensures a sufficient chiral discrimination  for obtaining and maintaining homochirality,
but provided that the water activity is sufficiently low \cite{Toxvaerd2018}. According to Pauling, the
$\alpha$-helix  structure is ensured by weak hydrogen-like bonds between units in
the helix. $\textit{But these bonds are in competition with hydrogen bonds to the attached wa-}$\\
$\textit{ter molecules at the protein, which destabilizes the   $\alpha$-helix structure at high}$ \\ 
$\textit{water activity (i.e. concentration).}$

It is, however, only the stability of homochirality in the proteins, which depends on the salinity of the aqueous cytosol
suspension,
because the homochirality of carbohydrates  is obtained and ensured  in a different way, by stereo
specific enzymes in the metabolism \cite{Toxvaerd2018}. All polymer units in carbohydrates
have a D-configuration at carbon atom No. 5 for hexose-units and
No, 4 for pentose-units. The carbohydrate polymers in biosystems are synthezised from
Glyceraldehyde-3-phosphate, which has a very active isomerization kinetics.
The isomerization kinetics for Glyceraldehyde-3-phosphate
is catalysed by the extreme effective enzyme, Triose Phosphate Isomerase,
and it looks like a paradox, that the bio-carbohydrate wold is 100\% homochiral,
when the key-molecule in the polymerization is chiral unstable. But
the explanation of this paradox is, that once the polymerization with Glyceraldehyde-3-phosphate 
starts by creation of a hexose or a pentose (Ribose), 
the chiral center in Glyseraldehyde-3-phosphate is preserved by stable covalent bonds
\cite{Toxvaerd2014}, and the total dominance of the D-configurations in 
bio-carbohydrates is obtained and maintained by {stereo specific enzymes  \cite{Toxvaerd2018}.

That proteins and enzymes are homochiral stable at a low water activity, is a daily life's observation.
One can conserve meat in a salty- or sugar solutions, whereby
the bacteria are dehydrated and inactive. But their life is only set on \ ``stand-by" in the salty solution, and some bacteria can even survive
in a dehydrated state under extreme conditions of pressure and temperature for a very long time \cite{Nicholson}.
It is therefore  natural to conclude, that the prebiotic
self-assembly of peptides  from units of amino acids  took place in an environment with higher salinity and lower water activity than in  the
cytosol, whereby one not only could obtain, but also maintain homochirality by the peptide synthesis \cite{Toxvaerd2017}.

\section{Darwin's  warm little pond}

 There has been many proposals to the geographic location(s),  where life occurred. The Hadean ocean(s),
 tidal pools, icy environments, mineral surfaces, alkaline hydrothermal vents, to mention some proposals,
 which also includes extraterrestrial places. 
This article focus on the necessity of a rather  salty solution with a relative high concentration of the building blocks of amino acids in order to
ensure a sufficient strength of chiral discrimination at the peptide polymerization. 
 It has always been a puzzle, from where the amino acids came, but in fact this question is irrelevant for  the start of the Abiogenesis.
 The problem is not from where they came, or whether or not the amino acids were in a racemic or homochiral composition.
The amino acids have an active isomerization kinetics, and an aqueous solution of amino acids  will racemize over time \cite{Bada}.
The problem is how to maintain a sufficient high concentration of amino acids for the polymerization,
and to how ensure a homochiral stable
form for a very long time. So long to that the other self-assembly of higher order structures and consecutive reactions
in bio systems could be established.
 The  necessity of
a high concentration of amino acids exclude the bulk ocean(s) and points to places with \ ``confine geometry".
Because, if the prebiodic polymerization of peptides
had occurred in a well stirred   Hadean ocean, it would require an enormous amount of amino acids. 

 In a  letter to a close friend  Darwin proposed, that life were started  \ `` ..in  some warm little pond.." \cite{Bada2009}, and since then
 there has been many suggestions to where in the aqueous environment the life processes started.
The Darwinism is often  presented as  a continuous coherent evolution from simple inorganic processes toward the
living systems. This outlook of the emergence of living systems 
 as prebiotic   \ `` chemistry in a bag" has  been criticised \cite{Branscomb}. Branscomb and Russell noticed, that all living systems exist in a self-generate  physical state, that is extremely
far from thermodynamic equilibrium i.e., equivalently, of extremely low entropy and thus of correspondingly low probability.
They  conclude,  that a simple mass-action chemistry cannot
explain the self-generated and far-from -equilibrium myriad of disequilibrium in a living organism.

The living organism is in a   \ `` far from equilibrium" state with gradients in the electrical potential, and protonic and ionic 
 concentrations.
These facts excludes most of the proposed locations for the prebiotic self-assembly
of the building blocks. The most favorable place, however, is the submarine alkaline hydrothermal vents,
where the confined environment with protonic, ionic and concentration  gradients fulfil
 the requirements for non-equilibrium self-assembly
of higher order organic structures \cite{Sousa,Branscomb,Russel,Sleep1}.
In this context it is interesting to notice, that the earliest sign of life is located
to be in  fluvial hot spring deposits \cite{Djokic2017}.

It is natural to connect the evolution of bio systems
with the geological evolution.
The  geology of the planet Earth has changed significantly over time, and the
physicochemical environment, where the prebiotic self-assembly took place,
must also have changed during the prebiotic time.
This fact might explain how the strong non-equilibrium chemistry with differences
in ion concentrations between the cytosol and the aqueous environment
have been established. Could it not have happened, that e.g.  emulsion of peptides were synthesised as \ ``chemistry
in a bag" in the vents or in the confine geometry  below the vents, and then were exposed to drastic changes in
concentrations caused be changes  in the vents?
LUCA is a space shuttle, not a shuttle in an empty space, but   \ `` chemistry in a bag" in a
hostile salty sea. We are used to consider the sodium pump as a physicochemical mechanism to ensure
a potential difference at the membrane and a gradient in sodium and potassium. But the ion channels also
ensure a low salinity in the cells, which is a 
necessity for our life processes.

\section{Discussion}
The Abiogenesis, or the origin of
life, is probably not a result of  a series of single events, but rather the result
of a gradual process with increasing complexity of molecules and chemical
reactions,
and the prebiotic synthesis 
  might not have left any trace  of the establishment
of   structures and reactions at the beginning of the evolution \cite{Hud,Fialho,Raiser}.
But the evolution have lead to different forms  of the most simple  living systems.
There exists two forms of simple procariotes: bacteria and the archaea, who mainly differ
with respect to  the constitutions of their membranes.
Their metabolism, genetics and enzyme systems are qualitative the same, and  they  are both homochiral
with respect to their peptide and carbohydrate units and with an interior of the cells, which is a diluted aqueous suspension.
These facts seems to indicate, that the prebiotic biosynthesis, the metabolism and a genetics were established to some extent,
before the life were established.
E.g. the central enzyme in Glycolysis/gluconeogenesis, Triose Phosphate
Isomerase may have been present in the proteome \cite{deFarias2016}. And the very fist
step in polymerization of carbohydrates in the metabolism, the synthesis
of D-fructose-1,6-bisphosphate from dihydroxyacetone phosphate and 
D-Glyceraldehyde-3-phosphate is controlled by a stereo specific enzyme,
Aldolase \cite{Munegumi2015}. An enzyme which also can be identified as an ancestral enzyme \cite{Say2010}. 

$\textit{The missing link}$  between a saline prebiotic environment and LUCA 
is the   advent of life:  a cell with a membrane, metabolism and genetics.
A cell, who although it was unstable, was self generating, 
whereby a wave of life were started.

One might object, that the hypothesis
about the emergence of homochirality in a  prebiotic saline aqueous environment
is  speculation, without a possibility of verification.
This is , however, not correct. Some
part of the hypothesis can be experimentally verified. It is straight forward
possible to test for homochiral stability of enzymes with respect to the  activity of water.
The prediction is that the homochiral stability  of  enzymes depends on the
degree of salinity of the aqueous suspension. One shall expect, that an enzyme
in pure water  is more chiral unstable than in an  $\textit{in vitro}$ cytosol solution, 
and that it is stable at a higher ionic concentration than in the cytosol.

The hypothesis is also,}
that the stability of homochirality in proteins is a necessity for obtaining and maintaining 
the homochirality of the carbohydrates \cite{Toxvaerd2018}. It might also be possible
to verify this part of the hypothesis: That the metabolism only acts with stereo specific enzymes,
whereby the homochirality of carbohydrates is ensured  by the kinetics in the metabolism \cite{Toxvaerd2018}.
The prediction is that also Ribose's metabolism is controlled by   
stereo specific enzymes, which ensures  a  D-Ribose wold, a necessity for RNA
and DNA. And that these enzymes are ancestral and present in the Progenote
and in LUCA.
If so, it establish an order in the evolution \cite{Toxvaerd2018}, and these behaviours together links the physicochemical state of the cytosol in the living cell together
with the instability of homochirality and the dead of a living organism in the global wave of life.

And it gives 
 an ideas of, where in the aqueous environment and how the prebiotic self-assembly took place.

				\section{Acknowledgment}
Jeppe C Dyre is gratefully acknowledged.
This work was supported by the VILLUM Foundation’s Matter project, grant No. 16515.


\begin{thebibliography}{00}
	\bibitem[Bada, 2004]{Bada2004} Bada, J. L., 2004. How life began on Earth: a status report.
		Earth and Planetary Science Letters 226, 1-15.
	\bibitem[lane et al., 2010]{Lane2010} Lane, N., Allen J.F., Martin W., 2010. How did LUCA make a living?
		Chemiosmosis in the origin of life. BioEssays 32, 271-280.
	\bibitem[Huber et al., 2012]{Huber} Huber, C., Kraus, F., Hanzlik, M., Eisenreich, W., W\"{a}chtersh\"{a}user, G., 2012.
		Elements of Metabolic Evolution. Chem. Eur. J. 18, 2063-2080.	
	\bibitem[Hud et al., 2013]{Hud} Hud, N. V., Cafferty, B. J., Krishnamurthy, R., Williams, L. D., 2013.
		The Origin of RNA and \ ``My Grandfather's Axe''. Chemistry \& Biology 20, 466-474.
	\bibitem[Pascal et al., 2013]{Pascal} Pascal, R., Pross, A., Sutherland, J. D., 2013.
		Towards an evolutionary theory of the origin of life based on kinetics and thermodynamics.
	 Open Biol 3, 130156.
 \bibitem [St\"{u}eken et al., 2013]{Stueken}  St\"{u}eken, E. E., Anderson, R. E., Bowman, J. S., Brazelton, W. J, Colangelo-Lillis, J.,
	 Goldman, A. D., Som, S. M., Baross, J. A., 2013.
	Did life originate from a global chemical reactor?  Geobiology 1, 101-126.
\bibitem[Sousa et al., 2013]{Sousa} Sousa, F. L., Thiergart, T., Landan, G., Nelson-Sathi, S., Pereira, I. A. C., Allen, J. F., Lane, N.,
	Martin, W. F., 2013.  Early bioenergetic evolution.  Phil. Trans. R Soc. B 368, 20130088.	
\bibitem[Sutherland, 2017]{Sutherland} Sutherland, J.D., 2017. Studies on the origin of
	life- the end of the beginning. Nature Reviews, Chemistry 1, 0012.	
\bibitem[Spitzer, 2017]{Spitzer} Spitzer, J., 2017. Emergence of Life on Earth: A Physicochemical Jigsaw Puzzle.
	J. Mol. Evol. 84, 1-7 (2017).
\bibitem[Walker, 2017]{Walker} Walker, S.I., 2017. Origins of life: a problem for physics,
	a key issues review. Rep. Prog. Phys. 80, 092601.	
\bibitem[Branscomb and Russell, 2018]{Branscomb} Branscomb, E., Russell, M.J., 2018.
	Frankenstein or a Submarine Alkaline Vent: Who Is Responsible for Abiogenesis? BioEssays 40, 1700179, 1700182.
\bibitem[Velasco, 2018]{Velasco2018} Velasco, J., 2018. Universal common ancestry, LUCA,
and the Tree of Life: three distinct hypotheses about the evolution of life. Biology \& Philosophy. 33, 1-18.
\bibitem[Higashi et al., 2018]{Higashi} Higashi, K., Kawai, Y., Baba, T., Kurokawa, K., Oshima, T., 2018.
	Essential cellular modules for the proliferation of the primitive cell.  Geoscience Frontiers 9, 1155-1161.


\bibitem[Woese and Fox, 1977]{WoeseFox1977}Woese, C. R., Fox. G. E., 1977. The Concept of Cellular Evolution.
J. Mol. Evol. 10, 1-6. 
\bibitem[Woese, 1987]{Woese1987}Woese, C. R., 1987. Bacterial Evolution. Microbiol. Rev. 51, 221-271.
\bibitem[Giulio, 2011]{DiGiulio2011}Giulio, M. D., 2011. The Last Universal Common Ancestor (LUCA)
and the Ancestors of Archaea and Bacteria were Progenotes. J. Mol. Evol. 72, 119-126.
\bibitem[Weiss et al., 2016]{Weiss2016}Weiss, M. C., Sousa, F. L., Mrnjavac, N.,
	Neukirchen, S., Roettger, M., Nelson-Sathi, S., Martin, W. F., 2016. The physiology and habitat of 
the last universal common ancestor. Nature Microbiol. 1, 1-8.
\bibitem[Giulio, 2018]{DiGiulio2018}Giulio, M. D., 2018. On Earth, there would be a number of fundamental
	kinds of primary cells - cellular domains - greater than or equal to four. J. Theor. Biol. 443, 10-17.
\bibitem[Weiss et al., 2018]{Weiss2018}Weiss, M. C., Preiner, M., Xavier, J. C., Zimorski, V. Martin, F., 2018.
The last universal common ancestor between ancient Earth chemistry and the onset of genetics. PLoS GENET. 14(8), e1007518.
\bibitem[Djokic et al., 2017]{Djokic2017}Djokic, T., Van Kranendonk, M. J., Campbell, K. A., Walter, M. R.,
Ward, C. R., 2017. Earlist signs of life on land preserved in ca. 3.5 Ga hot spring deposits. Nature Commun. 8, 15263.
\bibitem[Schopf et al., 2018]{Schopf2018}Schopf, J. W., Kitajima, K. Spicuzza, M. J., Kudryavtsev, A. B., Valley, 
	 J. W., 2018. SIMS analyses of the oldest known assemblage of microfossils
		document their taxon-correlated carbon isotope composition. PNAS 115, 53-58.
\bibitem[Tashiro et al., 2017]{Tashiro2017}Tashiro, T., Ishida, A., Hori, M., Igisu, M., Koike, M., M\'{e}jean, P., Takahata, N.,
 Sano, Y., Komiya, T., 2017. Early trace of life from 3.95 Ga sedimentary rocks in Labrador, Canada. Nature 549, 516-518.
\bibitem[Retallack and Noffke, 2019]{Retallack2019}Retallack, G. J., Noffke, N., 2019.
	Are there ancient soils in the 3.7 Ga
Isua Greenstone Belt, Greenland? Palaeogeogr., Palaeoclimatol., Palaeoecol. 514, 18-30.


\bibitem[Harrison, 2009]{Harrison} Harrison, T.M., 2009. The Hadean Crust:
	Evidence from  $>$4 Ga Zircons. Annu. Rev.  Earth  Planet. Sci.  37, 479-505.
\bibitem[Sleep, 2010]{Sleep} Sleep, N.H., 2010. 	 The Hadean-Achaean Environment. Cold Spring Harb Perspect. Biol. 2: a002527.
	
	\bibitem[Martinez, 1998]{Martinez1998}Mart\'{i}nez, D. E., 1998. Mortality patterns suggest lack of senescence in 
	Hydra. Exp Gerontol. 33, 217-225.
	\bibitem[He et al, 2015]{He2015}He, J. H., Zheng, L., Zhang, W., Lin, Y., 2015. Life Cycle Reversal in
		\textit{Aurelia} sp. 1(Cnidaria Scyphozoa). PLoS ONE 10(12, e0145314.
	\bibitem[Cong et al., 2002]{Cong2002}Cong, Y-S., Wright, W. E., Shay, J. W., 2002. Human Telomerase and Its
Regulation. Microbiol Mol. Biol. Rev. 66, 407-425. 
	\bibitem[Shepherd, 2006]{Shepherd2006}Shepherd, V.A., 2006. The Cytomatrix as a Cooperative System of
		Macromolecular and Water Networks. Curr. Top. Dev. Biol 75, 171-223.
\bibitem[Norris et al., 2007]{Norris2007}Norris, V., den Blaauwen, T., Cabin-Flaman, A., Doi, R. H., Harshey, R.,
	Janniere, L., Jimenez-Sanchez, A., Jin, D. J., Levin, P. A., Mileykovskaya,
	E. Minsky, A., Saier, Jr., M., Skarstad, K., 2007. Functional Taxonomy
	of Bacterial Hyperstructures. Microbiol. Mol. Biol. Rev. 71, 230-253.
\bibitem[Luby-Phelps, 2000]{Luby-Phelps2000}Luby-Phelps, K., 2000. Cytoarchitecture and Physical Properties of
	Cytoplasm: Volume, Viscosity, Diffusion, Intracellular Surface Area.  Int.
	Rev. Cytol. 192, 189-221.
\bibitem[Maruyama et al., 2013]{Maruyama} Maruyama, S., Ikoma, M., Genda, H., Hirose, K., Yokoyama, T., Santosh, M., 2013.
	The naked planet Earth: Most essential pre-requisite for the origin and evolution of life. Geoscience Frontiers   4, 141-165.		
\bibitem[Aoki et al., 2018]{Aoki} Aoki, S., Kabashima, C., Kato, Y., Hirata, T., Komiya, T., 2018. Influence of  contamination
	 on banded iron formations in the Isua supracrustal belt, West Greenland: Reevaluation of the Eoarchean seawater compositions.
		Geoscience Frontiers 9, 1049-1072.
	\bibitem[Marty et al., 2018]{Marty} Marty, B., Avice, G., Bekaert, D.V., Broadley, M.W., 2018.
	 Salinity of the Achaean oceans from analysis of fluid inclusions in quartz. C. R. Geo science  350, 154-163.
\bibitem[Bada, 1972]{Bada}  Bada, J. L., 1972.  Kinetics of Racemization of Amino Acids as a Function of pH.
	J. Am. Chem. Soc. 94, 1371-1373 (1972).
\bibitem[Fuji et al., 2018]{Fujii2018} Fujii, N., Takata, T., Fujii, N., Aki, K., Sakaue, H., 2018. D-Amino acids in protein: The mirror
	of life as a molecular index of aging. BBA-Proteins and Proteomics 1866,  840-847.
		1,6-bisphosphate aldolase/phosphatase may be an ancestral gluconeogenic enzyme. Nature 464, 1077-1081.
\bibitem[Toxvaerd, 2017]{Toxvaerd2017} 
	Toxvaerd, S., 2017. The role of the peptides at the origin of life. J. Theor. Biol. 429, 164-169.		
\bibitem[Pauling et al., 1951]{Pauling1} Pauling, L., Corey, R. B., Branson, H. R., 1951. The structure of proteins:
	Two hydrogen-bonded helical configurations of the polypeptide chain. Proc. Natl. Acad. Sci. 37, 205-211.
\bibitem[Pauling and Corey, 1951]{Pauling2} Pauling, L., Corey, R.B., 1951. 
	The pleated sheet, a new layer configuration of polypeptide chains. Proc. Natl. Acad. Sci, 37, 251-256.
\bibitem[Toxvaerd, 2018]{Toxvaerd2018} Toxvaerd, S., 2018.The start of the Abiogenesis: Preservation of homochirality
	in proteins as a necessary and sufficient
	condition for the establishment of the metabolism.  J. Theor. Biol. 451, 117-121.
\bibitem[Toxvaerd, 2014]{Toxvaerd2014} Toxvaerd, S., 2014. The role of carcohydrates at the origin of homochirality in biosystems.
	Orig. Life. Evol. Biosph. 44, 391-409.
\bibitem[Nicholson et al., 2000]{Nicholson} Nicholson, W.L., Munakata, N., Horneck, G., Melosh, H.J., Setlow, P., 2000.
	Resistance of $\textit{Bacillus}$ Endospores to Extreme Terrestrial and Extraterrestrial Environments.
		Microbiol. Mol. Biol. Rev.  64, 548-572.
	\bibitem[Peret\'{o} et al., 2009]{Bada2009} Peret\'{o} J., Bada J.L., Lazcano A., 2009. Charles Darwin and the Origin of Life.
		Orig. Life. Evol. Biosph.  39, 395-406.
	\bibitem[Russell et al., 1994]{Russel} Russell, M. J., Daniel, R. M., Hall, A. J., Sherringham, J. A., 1994. A hydrothermally  precipitated
		catalytic iron sulphide membrane as a first step toward life. J. Mol. Evol. 39,  231-243.
	\bibitem[Sleep et al., 2011]{Sleep1} Sleep, N. H., Bird, D. K., Pope, E. C., 2011.
		Serpentinite and the dawn of life. Phil. Trans. R. Soc. B 366,  2857-2869.
	\bibitem[Fialho et al., 2018]{Fialho} Fialho, D. M., Clarke, K. C., Moore, M. K., Schuster, G. B., Krishnamurthy, R., Hud, N. V., 2018.
		Glycosylation of a model proto-RNA nucleobase with non-ribose sugars: implications
		for the prebiotic synthesis of nucleotides. Org. Biomol. Chem.  16,  1263-1271.
	\bibitem[Raiser, 2018]{Raiser} Raiser, M., 2018. An appeal to magic?
		The discovery of a non-enzymatic metabolism and its role in the origins of life. Biochemical J. 475, 2577-2592.
	\bibitem[de Farias  et al., 2016]{deFarias2016}de Farias, S. T., R\^{e}go, T. G., Jos\'{e}, M., 2016.
A proposal of the proteome before the last universal common ancestor
		(LUCA). Intern. J. Astrobiol. 15, 27-31.
	\bibitem[Munegumi, 2015]{Munegumi2015}Munegumi, T., 2015. Aldolase as a Chirality Intersection of L-amino
		Acids and D-Sugars. Orig. Life. Evol. Biosph. 45, 173-182.
	\bibitem[Say and Fuchs, 2010]{Say2010}Say, R. F., Fuchs, G., 2010. Fructose
		1,6-bisphosphate aldolase/phosphatase may be an ancestral gluconeogenic enzyme. Nature 464, 1077-1081.
\end{thebibliography}
\end{document}